\documentclass[times]{simauth}

\usepackage{xr}
\usepackage{moreverb}
\usepackage{graphicx}

\usepackage[
colorlinks,bookmarksopen,bookmarksnumbered,citecolor=red,urlcolor=red]{hyperref}

\newcommand\BibTeX{{\rmfamily B\kern-.05em \textsc{i\kern-.025em b}\kern-.08em
T\kern-.1667em\lower.7ex\hbox{E}\kern-.125emX}}

\def\volumeyear{2019}

\usepackage[font={bf, scriptsize}]{subcaption}
\usepackage{xspace}
\usepackage{array}

\begin{document}

\runninghead{L. Li, T. Wu, and C. Feng}

\title{Model Diagnostics for Censored Regression via Randomized Survival Probabilities}

\author{Longhai Li\corrauth\footnotemark[2], Tingxuan Wu\footnotemark[2]\footnotemark[3],  and Cindy Feng\footnotemark[3]}

\address{\footnotemark[2] Department of Mathematics and Statistics, University of Saskatchewan, 106 Wiggins Rd, Saskatoon, SK, S7N5E6, Canada. 
\\ \footnotemark[3] School of Public Health, University of Saskatchewan, 104 Clinic Place, Saskatoon, SK, S7N5E5 Canada.}
\corraddr{longhai@math.usask.ca}

\begin{abstract}

Residuals in normal regression are used to assess a model's goodness-of-fit (GOF) and discover directions for improving the model. However, there is a lack of residuals with a characterized reference distribution for censored regression. In this paper, we propose to diagnose censored regression with normalized randomized survival probabilities (RSP). The key idea of RSP is to replace the survival probability of a censored failure time with a uniform random number between 0 and the survival probability of the censored time. We  prove that RSPs always have the uniform distribution on $(0,1)$ under the true model with the true generating parameters. Therefore, we can transform RSPs into normally-distributed residuals with the normal quantile function. We call such residuals by normalized RSP (NRSP residuals). We conduct simulation studies to investigate the sizes and powers of statistical tests based on NRSP residuals in detecting the incorrect choice of distribution family and non-linear effect in covariates.  Our simulation studies show that, although the GOF tests with NRSP residuals are not as powerful as a traditional GOF test method, a non-linear test based on NRSP residuals has significantly higher power in detecting non-linearity.  We  also compared these model diagnostics methods with a breast-cancer recurrent-free time dataset.  The results show that the NRSP residual diagnostics successfully captures a subtle non-linear relationship in the dataset, which is not detected by the graphical diagnostics with CS residuals and existing GOF tests. 

\end{abstract}

\keywords{goodness-of-fit, residual diagnostics, Cox-Snell residual, quantile residual, model checking}

\maketitle

\footnotetext[4]{\textbf{List of Abbreviations}: AIC, Alkaike's Information Criterion; AFT, accelerated failure time;  CHF, cumulative hazard function; CS, Cox-Snell; GOF, goodness-of-fit; KM, Kaplan-Meier; KS,  Kolmogorov-Smirnov; LCKS, Lilliefors-Corrected Kolmogorov-Smirnov; SW, Shapiro-Wilk; SF, Shapiro-Francia; SP, survival probability; USP, unmodified SP; RSP, randomized SP; MSP, modified SP; NRSP, normalized RSP; NMSP, normalized MSP; NUSP, normalized USP.}


\def \tistar {t^{*}_{i}}
\def \Tistar {T^{*}_{i}}
\def \SiTistar {S_i(\Tistar)}
\def \rcTistar {r^c_i (\Tistar)}
\def \SP{survival probability\xspace}
\def \SPs{survival probabilities\xspace}
\def\nmsp{\mbox{\scriptsize NMSP}}
\def\nrsp{\mbox{\scriptsize NRSP}}
\def\rsp{S_i^{R}(T_i, d_{i}, U_{i})}
\def \nref {[\textbf{references}]\xspace}
\def\sp{S_{i}(T_{i}^{*})}
\def\svf{S_{i}(\cdot)}
\def\ques{\textbf{[questions]}}
\def\pmin{p_{\mbox{\scriptsize min}}}
\def\obs{^{\scriptsize \mbox{obs}}}

\section{Introduction}
Model diagnostics is a crucial step in model building to ensure the validity of the statistical inference. Residual analysis is a conventional tool for model checking and diagnostics. Residuals of a model are used to check the overall goodness-of-fit (GOF) of a model, discover the direction for improving the model, and identify outlier observations. Cox-Snell (CS) residuals \cite{cox_general_1968} are widely used for checking survival regression models for failure times. CS residuals are transformed from the survival probabilities with the quantile function of the exponential distribution. When failure times are not censored, and the postulated model is the true model for them, the survival probabilities are uniformly distributed; hence, CS residuals are exponentially distributed. This reference distribution is the basis for model checking with CS residuals. The cumulative hazard function (CHF) of CS residuals is commonly plotted and compared to the $45^{\circ}$ straight line (unity slope and zero intercept). We can also employ some goodness-of-fit (GOF) tests such as the Kolmogorov-Smirnov (KS) test \cite{massey_kolmogorov-smirnov_1951} to check the exponentiality of CS residuals. Unfortunately, when there exist censored failure times, the survival probabilities are no longer uniformly distributed. Correspondingly, CS residuals are no longer exponentially distributed. Indeed, CS residuals are typically randomly censored observations from a distribution. We can still estimate the survival function of CS residuals  with KM-like methods  \cite{kaplan_nonparametric_1958,hirsch_plotting_1987} that can consider censoring and compare the CHF of censored CS residuals against the $45^{\circ}$ straight line for model checking. This plot is probably the most commonly used method in practice. Although it is not conducted very often in practice, the agreement of an empirical distribution with a reference distribution can also be quantified by some GOF test methods that are extended to handle censored data \cite{verrill_asymptotic_1987, verrill_tables_1988, royston_toolkit_1993, koziol_cramer-von_1976, csorgo_koziol-green_1981, chen_correlation_1984, akritas_pearson-type_1988, hollander_chi-squared_1992, cao_bayesian_2010, kim_tests_2019, millard_envstats_2018-1, steven_p_millard_author_envstats_2013}. 

The GOF checking that assesses the agreement of the distribution of residuals with a reference as described above is just the first-line model diagnostics, like checking patients' blood pressure in medical diagnostics.  The departure of the distribution of residuals of a flawed model may be too subtle to be detected by GOF checking.  More importantly, when a model failure is detected, the GOF test results typically reveal little information about the nature of the model failure, such as non-linear covariate effect, non-constant variances, and lack of independence. For identifying these model discrepancies, we need to conduct more in-depth graphical and quantitative diagnostics. Therefore,  the GOF checking is insufficient for practical model diagnostics. We also desire to define residuals that have a known reference distribution under the true model so that we can freely conduct a wide variety of model diagnostics. Some non-random methods have been proposed to modify CS residuals \cite{collett_modelling_2015}, which include the ways of adding CS residuals for censored times with a constant, the martingale residuals \cite{therneau_martingale-based_1990}, the deviance residuals, and possibly others \cite{therneau_modeling_2013}. However, these modified residuals under the true model do not have a unified and characterizable distribution. 

This paper proposes using normalized randomized survival probabilities (RSPs) to conduct model diagnostics for censored regression. The key idea of RSP is to replace the survival probability of a censored failure time with a uniform random number between 0 and the survival probability of the censored time. We prove that RSPs always have the uniform distribution on $(0,1)$ under the true model. Therefore, we can transform RSPs into normally-distributed residuals with the normal quantile function. We call such residuals by \textbf{normalized RSP (NRSP) residuals}. We conduct simulation studies to investigate the sizes and powers of statistical tests based on NRSP residuals in detecting the incorrect choice of distribution family and non-linear effect in covariates.  Our simulation studies show that the sizes of model diagnostic tests based on NRSP residuals are very close to nominal.  Furthermore, our results show that, although the GOF tests with NRSP residuals are not as powerful as a traditional GOF test method, a non-linearity test based on NRSP residuals has significantly higher powers in detecting the non-linear covariate effects.  We also compared these model diagnostic methods with a breast-cancer recurrent-free time dataset.  The results show that the model diagnostics with the NRSP residuals successfully captures a non-linear covariate effect in the dataset,  which is not detected by the graphical diagnostics with CS residuals and existing GOF tests. 

The rest of this paper is organized as follows. Section \ref{sec:review} reviews the existing residuals and diagnostic methods for censored regression.  In Section \ref{sec:rsp}, we present the definition of randomized survival probabilities (RSPs), with an illustrative example and proof of the uniformity of RSPs under the true model. In Section \ref{sec:sim}, we conduct simulation studies to investigate the performances of NRSP residuals.  Section \ref{sec:realdata} presents the results of applying the NRSP residual to  a breast cancer recurrence-free time dataset. The article is concluded in Section \ref{sec:conc}.  

\section{Review of Existing Model Diagnostics Methods for Censored Regression}\label{sec:review}

In this section, we review some existing model diagnostic methods used in survival analysis. A central concept in these residuals is the survival probability (SP). Suppose $T_{i}^{*}$ is the true failure time of the $i$th individual, which we assume to be a continuous random variable in this article. Let $t_{i}^{*}$ denote the realization of $T_{i}^{*}$.  In many practical problems, we may not be able to observe $t_{i}^{*}$ exactly, but we can observe that $T_{i}^{*}$ is greater than a value $C_{i}$, which is called right-censoring.  The observed failure times are denoted by the pair $(T_i, d_{i})$, where $T_{i}=\min(\Tistar, C_i), d_i=I (\Tistar < C_i)$. Since we will consider only the right-censoring in this article, we will use the ``censoring'' as a short for the ``right-censoring''.  

Suppose the survival function of $\Tistar$ based on a postulated model is defined as $S_{i}(\tistar) = P(\Tistar > \tistar)$, where the subscript $i$ indicates that the probability depends covariate $x_{i}$ for the $i$th individual. Using a simple probability argument, one can prove that the survival probabilities $S_i(\Tistar)$ are uniformly distributed when $\svf$ is the survival function of the true model for $\Tistar$.  
SPs can be transformed into random variables with a desired distribution by applying its inverse CDF or survival function. The widely used Cox-Snell (CS) residual is defined as $r^c_i (\Tistar)=- \log({S_i}(T_{i}^{*}))$,
where $-\log(\cdot)$ is the inverse survival function of  $\exp(1)$. Therefore, CS residuals are exponentially distributed under the true model. Although it is not used often in practice,  one can also define normally-distributed residuals  \cite{nardi_new_1999-1}:  $r^{n}_{i}(\Tistar)=\Phi^{-1}(\SiTistar)$, which we call by \textbf{normalized SPs}. Then we can apply a variety of residual diagnostic methods for normal regression to diagnose $\svf$. 

If $T_{i}$ is censored,  the \textbf{unmodified survival probability (USP)}, $S_{i}(T_{i})$, is larger than $S_{i}(\Tistar)$ since $T_{i}<\Tistar$.  Thus, when there are censored observations, the distribution of $S_{i}(T_{i})$ is no longer uniformly distributed under the true model.  The non-uniformity in USPs causes the difficulty of performing residual diagnostics. The \textbf{unmodified CS residuals}, $r_{i}^{c}(T_{i})=-\log(S_{i}(T_{i}))$, and \textbf{normalized unmodified  SPs (NUSP)}, $r^{n}_{i}(T_{i})$,  can be treated as univariate data with censoring if we ignore $x_{i}$.  In practice, the most widely used diagnostic tool is to apply KM methods \cite{kaplan_nonparametric_1958} to $\{(r_{i}^{c}(T_{i}), d_{i})| i=1,\ldots, n\}$ to get an estimate of the CHF of  CS residuals.  Under the true model,  the CHF of CS residuals is expected to be close to  the $45^{\circ}$ straight line.  In addition to the graphical checking,  we also desire a quantitative measure of the GOF.  This problem becomes challenging due to censoring. However, some methods have been developed for checking the GOF of univariate data with censoring.   Shapiro-Wilk (SW) and Shapiro-Francia (SF) normality tests \cite{shapiro_analysis_1965,shapiro_approximate_1972} have been extended to singly censored data \cite{verrill_asymptotic_1987, verrill_tables_1988, royston_toolkit_1993}.  Although censoring times $C_{i}$ for $\Tistar$ may not be random,  unmodified SPs and their transformations are typically randomly censored due to \textit{the randomness in covariates}.   The chi-squared test and some normality tests have been extended to randomly censored data; see \cite{koziol_cramer-von_1976, csorgo_koziol-green_1981, chen_correlation_1984, akritas_pearson-type_1988, hollander_chi-squared_1992, cao_bayesian_2010, kim_tests_2019} among others.  The function \texttt{gofTestCensored} in R package \texttt{EnvStats}  \cite{millard_envstats_2018-1, steven_p_millard_author_envstats_2013} provides an SF test  for multiply censored data. The method used in \texttt{gofTestCensored} is a generalization of the method for  extending SW and SF tests  for singly censored data. The key idea of these extensions is to measure the product-moment correlation between the uncensored observations and the corresponding standard normal quantiles with the linking probabilities (called plotting positions) estimated with KM-like methods \cite{kaplan_nonparametric_1958, hirsch_plotting_1987}. The details are given in the \href{https://www.rdocumentation.org/packages/EnvStats/versions/2.3.1/topics/gofTestCensored}{manual page} of the function \texttt{gofTestCensored}, which also contains a detailed discussion of SW and SF tests.

The graphical and quantitative methods for comparing the distribution of residuals to a reference distribution are useful in detecting the lack of model fit. Many model mis-specifications may be captured by these methods.  However,  examining the distribution of residuals alone is only the first-line model diagnostics. An analogy in medical diagnostics is that we check the blood pressure or temperature of patients for detecting diseases, which is not sufficiently powerful for the identification of potential diseases.  For example, these methods ignore the covariates. Thus, they may fail to detect the model mis-specification in linking $T_{i}^{*}$ with $x_{i}$.  More importantly, the GOF test results typically cannot reveal the nature of model mis-specification, especially that related to $x_{i}$, such as non-linearity, lack of independence, non-constant variances, and many others. For conducting such in-depth diagnostics, we desire to define residuals  that have a known reference distribution under the true model.  A number of methods have been proposed to modify USPs or their transformations. A commonly used method is to shrink the USPs of the censored failure times: 
\begin{equation}
S_i^{\prime}(T_i, d_{i}, \eta) =
\left\{
\begin{array}{rl}
S_{i}(T_{i}), & \text{if $T_i$ is uncensored, i.e., $d_{i}=1$,}\\
\eta \, S_{i}(T_{i}), & \text{if $T_i$ is censored, i.e., $d_i=0$,} 
\end{array}
\right. 
\end{equation}
where  $\eta\in(0,1)$. We call the shrunken SPs by \textbf{modified SPs (MSPs)}. Transforming the MSPs with the inverse survival of $\exp(1)$ results in the modified CS residuals with a constant $\Delta=-\log(\eta)$ added to the CS residuals of censored failure times, given by ${r_i^c}^{\prime}(T_{i}, d_{i}, \Delta) =-\log(S_i^{\prime}(T_i, d_{i}, \eta)))=r^c_i (T_{i}) +\Delta (1-d_{i})$. We can similarly define \textbf{normalized MSPs (NMSP)}  \cite{nardi_new_1999-1}: ${r_i^{\nmsp}}(T_{i}, d_{i},\eta) = \Phi^{-1} (S_i^{\prime}(T_i, d_{i}, \eta))$.
There are many different choices for the shrinkage factor $\eta$ or $\Delta$ in the literature based on different choices of conditional means of SPs or their transformations given $T_{i}^{*}>T_{i}$; $\Delta = 1$ and $\Delta=\log(2)$ are often chosen; see \cite{collett_modelling_2015, therneau_modeling_2013,  nardi_new_1999-1}. Other residuals, for example the martingale residuals $r^{M}_{i}(T_{i}, d_{i})=d_{i}-r^{c}_{i}(T_{i})$ and the deviance residuals, and residual-based diagnostic tools have also been proposed for diagnosing censored regression; see \cite{collett_modelling_2015, therneau_martingale-based_1990, therneau_modeling_2013,residual_mixturecuremodel_2017,  grambsch_proportional_1994, residual_Tree-Structured_survival, residual_ph_interval_censored, deviance_normal_scores, lin_checking_1993, law_residual_2017,shepherd_probability-scale_2016-1, hillis_residual_1995} and the references therein.   Although many residuals by modifying the USPs or their transformations exist, a common drawback for these modified residuals is that their distributions under the true model are very complicated due to censoring, thus, they cannot be characterized clearly with a known distribution or probability table. This distribution depends on the distribution of censoring times $C_{i}$, which varies for different datasets. Therefore, there is a lack of reference distributions for us to conduct model diagnostics with these residuals.

\section{Normalized Randomized Survival Probabilities}\label{sec:rsp}
\subsection{Definition of Randomized Survival Probabilities}
We will propose to diagnose censored regression with normalized randomized survival probabilities (RSPs).  The randomized survival probability (\textbf{RSP}) for $T_{i}$ is defined as:
\begin{equation}
S_i^{R}(T_i, d_{i}, U_{i}) =
\left\{
\begin{array}{rl}
S_{i}(T_{i}), & \text{if $T_i$ is uncensored, i.e., $d_{i}=1$,}\\
U_{i}\,S_{i}(T_{i}), & \text{if $T_i$ is censored, i.e., $d_i=0$,} 
\end{array}
\right. \label{rsp} 
\end{equation}
where $U_i$ is a uniform random number on $(0, 1)$, and $S_{i}(\cdot)$ is the postulated survival function for $\Tistar$ given $x_{i}$. From the definition, we see that  the fixed shrinkage factor $\eta$ in MSPs is replaced by a random number $U_{i}\in (0,1)$. In other words,  $S_i^{R}(T_i, d_{i}, U_{i})$ is a random number between $0$ and $S_{i}(T_{i})$ when $T_{i}$ is censored.  We will show that the randomized SP is uniformly distributed on $(0,1)$ given $x_{i}$ under the true model. Therefore, we can transform them into residuals with any desired distribution. We prefer to transforming them with the normal quantile:
\begin{equation}
r_{i}^{\nrsp}(T_i, d_{i}, U_i)=\Phi^{-1} (S_{i}^R(T_i, d_{i}, U_i)).\label{nrsp}
\end{equation}
We name the residuals in \eqref{nrsp} as normalized randomized SP ( \textbf{NRSP}) residuals. Due to the normality of NRSP residuals under the true model, we can conduct model diagnostics with NRSP residuals for censored data in the same way as conducting model diagnostics for a normal regression. There are a few advantages of transforming RSPs into NRSPs. First, the diagnostic methods for checking normal regression are rich in the literature.  Second, transforming RSPs into normal deviates facilitates the identification of extremely small and large RSPs. The frequency of such small RSPs may be too small to be highlighted by plotting RSPs. However, the presence of such extreme SPs, even very few, is indicative of model mis-specification. Normal transformation can highlight such extreme RSPs.
\subsection{Illustration of the Uniformity of RSP}

\begin{figure}[htp]
  \centering
  \begin{subfigure}{0.245\textwidth}
    \centering
    \includegraphics[width=\textwidth, height=1.5in]{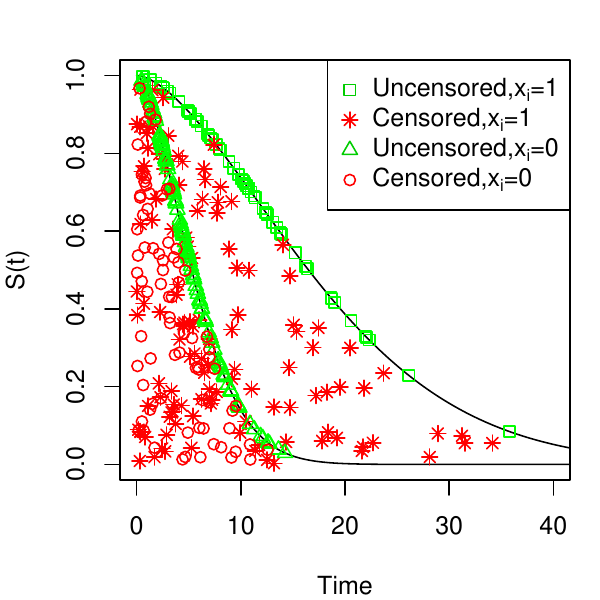}
    \caption{RSPs, True Model\label{fig:rsp_t}}
  \end{subfigure}   
  \begin{subfigure}{0.245\textwidth}
    \centering
    \includegraphics[width=\textwidth, height=1.5in]{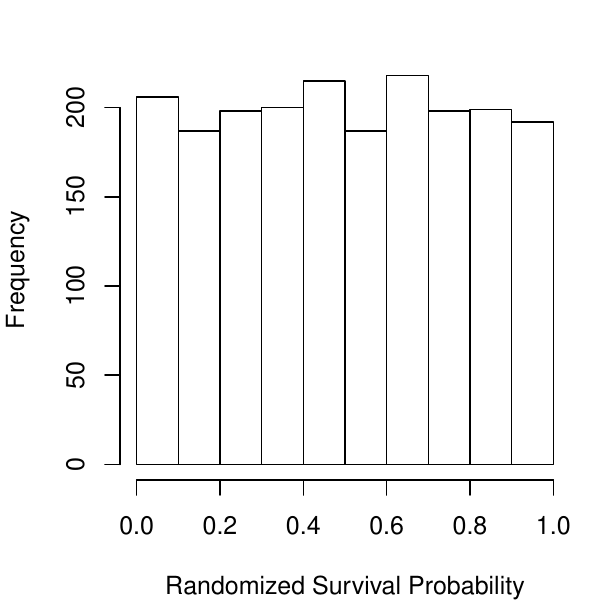}
    \caption{Histogram of RSPs, True Model\label{fig:his_rsp_t}}
  \end{subfigure}  
   \hfill
   \begin{subfigure}{0.245\textwidth}
    \centering
    \includegraphics[width=\textwidth, height=1.5in]{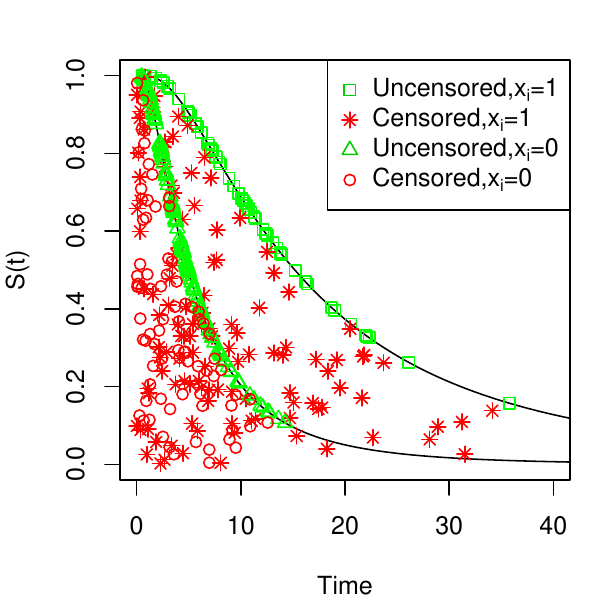}
     \caption{RSPs, Wrong Model\label{fig:rsp_w}}
  \end{subfigure} 
   \hfill
  \begin{subfigure}{0.245\textwidth}
    \centering
    
    \includegraphics[width=\textwidth, height=1.5in]{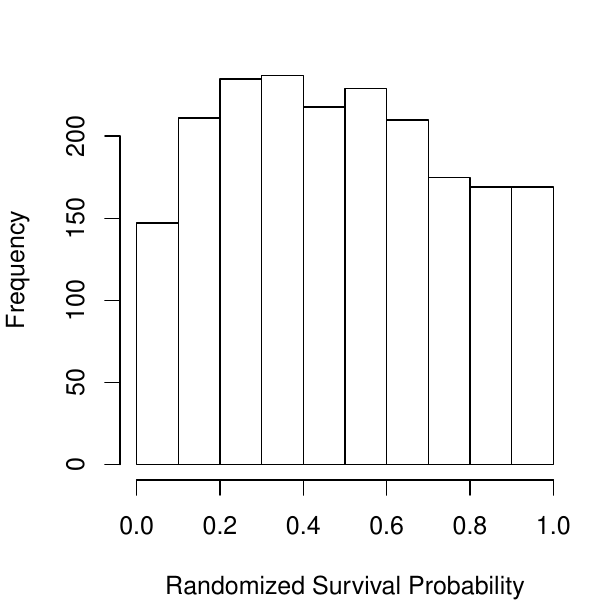}
    \caption{Histogram of RSPs, Wrong Model\label{fig:his_rsp_w}}
  \end{subfigure}  
  
\caption{Illustration of the uniformity of RSPs.  An animated display of this figure with multiple simulated datasets is shown in  the URL given in Section \ref{sec:animation}. \label{fig:rsp}}
 \end{figure}

We generate $2000$ failure times, $\Tistar$, as follows: $\log(T_{i}^{*})= 2 + x_{i} + \epsilon_i$, where $\epsilon_{i}$ is generated from the extreme-value distribution with a shape parameter $\gamma$ set as 1.8, and $x_{i}$ is generated as a Bernoulli (0.5). This model is called Weibull accelerated failure time (Weibull AFT) model \cite{collett_modelling_2015, george_survival_2014,weibull_model_analysis_survivaldata}.  Figure \ref{fig:rsp_t} depicts 400 RSPs along with the two fitted survival curves when Weibull model is fitted to the dataset. For the uncensored times, the survival probabilities are calculated with the survival functions and for each censored time $T_{i}$, the survival probability evaluated at observed $T_{i}$ is replaced by a random number between 0 and $S_{i}(T_{i})$.  The histogram in Fig. \ref{fig:his_rsp_t} shows clearly that the RSPs are uniformly distributed on $(0,1)$.  We also fitted log-normal model as a wrong model for this dataset with results shown by Figure \ref{fig:rsp_w} and \ref{fig:his_rsp_w}. We see that due to the mis-specified distribution family, the RSPs are not uniformly distributed.

\subsection{Proof of the Uniformity of RSP} \label{sec:proof}

We first rewrite $\rsp$ as a function of ($T_{i}^{*}, C_{i}, U_{i}$) as follows:
\begin{equation}
S_i^{R'}(T_i^{*}, C_{i}, U_{i}) =
\left\{
\begin{array}{rl}
S_{i}(T_{i}^{*}), & \text{if } T_i^{*} \leq C_{i}\\
U_{i}\,S_{i}(C_{i}), & \text{if } T_i^{*}>C_{i}.
\end{array}
\right. \label{rsp2} 
\end{equation}
We will show that the conditional distribution of $S_i^{R'}(T_i^{*}, C_{i}, U_{i})$ given $C_{i}=c$ is uniform on $(0,1)$. \textit{To proceed, we assume that $\Tistar$ and $C_{i}$ are independent, that is, censoring times are non-informative to the original failure times.}  Based on this assumption, the distribution of $S_{i}(\Tistar)$ is untacted given $C_{i}=c$. Hence, given $\Tistar \leq c$, the RSP is equal to $S_{i}(\Tistar)$, and it is uniformly distributed on $(S_{i}(c), 1)$. And, given $\Tistar > c$, the RSP is equal to $U_{i}\,S_{i}(c)$, which is uniformly distributed on $(0, S_{i}(c))$ due to the uniformity of $U_{i}$. In addition, the probability of $\Tistar >c$ given $C_{i}=c$ is $S_{i}(c)$. With $\lambda(B)$ denoting the length of an interval $B$ on $(0,1)$,  we can derive the following equations:
\begin{eqnarray}
\lefteqn{P(S_i^{R'}(T_i^{*}, C_{i}, U_{i})\in B\ |\ C_{i}=c)} &&\\
&= & P(S_i(T_i^{*})\in B\ |\ C_{i}=c, \Tistar \leq c)\times P(\Tistar \leq c) + P(U_{i}S_{i}(c)\in B\ |\ C_{i}=c, \Tistar > c)\times P(\Tistar > c) \\
&=& \lambda (B\cap (S_{i}(c), 1)) + \lambda (B\cap (0, S_{i}(c))) = \lambda (B)
\end{eqnarray}
Since the conditional distribution of $S_i^{R'}(T_i^{*}, C_{i}, U_{i})$ given $C_{i}=c$ is uniform on $(0,1)$, the marginal distribution of $S_i^{R'}(T_i^{*}, C_{i}, U_{i})$ is uniform on $(0,1)$ too after the $C_{i}$ is marginalized away by applying the total probability rule again. The proof that the randomized SPs are uniformly distributed on $(0,1)$ is completed. 

It is worth pointing out that we have proved that the RSP given $x_{i}$ is uniformly distributed on $(0,1)$.  Therefore, the marginal distribution of RSPs is also uniformly distributed on $(0,1)$. The marginal uniformity is used in GOF tests, and the conditional uniformity of RSPs can be used to check model assumptions in linking $\Tistar$ with $x_{i}$. 

\subsection{Model Diagnostics Based on NRSP Residuals}

NRSP residuals can be used in testing a model's GOF,  for which we recommend SW or SF normality tests. In the proof given in Section \ref{sec:proof}, we assume that the postulated model $\svf$ is the true model for $\Tistar$. In practice, the $\svf$ needs to be estimated with sample data. When the same dataset, $\{(T_{i},d_{i})|i = 1,\ldots, n\}$, is used to estimate the model parameters and used to calculate residuals for model checking, there might be a conservatism (bias) problem due to the double use of the dataset. NRSP residuals may be more concentrated around 0 than exactly distributed as $N(0,1)$. However, this conservatism is very small when the sample size $n$ is much larger than the number of parameters. For GOF tests, our simulation results show that the SW and SF normality tests applied to NRSP residuals are more resistant to the conservatism than the KS test; see a dedicated investigation in Section \ref{sec:ks} with simulation studies.  However, when a very complex model (for example, with many covariates) is fitted to a small number of failure times, it may be necessary to apply cross-validation methods to compute NRSP residuals.  For example, in leave-one-out cross-validation, an observation is held out; then, the model parameters are estimated with the remaining observations; finally, the estimated parameters are used to calculate the residual for the held-out observation.  

When a model is correctly specified, the conditional distribution of NRSPs given $x_{i}$ is approximately standard normal and is homogeneous for varying $x_{i}$ and the linear predictors (fitted values). Therefore, most model diagnostic tools for normal regression can be applied to NRSPs for diagnosing censored regression. In particular, a scatterplot of NRSP residuals versus each $x_{i}$  can be used to check whether the linear assumption with $x_{i}$ is appropriate or not. For qualitatively testing the non-linearity, we apply the $F$-test in ANOVA for testing the equality of means of NRSPs in the $k$ groups that are formed by cutting fitted values with equally-spaced intervals. 

\subsection{A P-value Upper Bound for Assessing Replicated NRSP GOF Test p-values}\label{sec:pmin}

A difficulty in conducting statistical tests with NRSP residuals is the randomness in the test p-values.  Given a fitted model, we can generate multiple sets of NRSP residuals and obtain replicated test p-values. We can choose to randomly report one of them and draw a histogram of the replicated test p-values to assess the model fit. However, the randomness may still be undesirable. In this section, we describe a theoretically non-random p-value upper bound based on a formula about the distribution of order statistics of correlated random variables.  Suppose $d_{1},\ldots,d_{J}$ are possibly correlated random variables with a common survival function $S(\cdot)$. Let $d_{(r)}$ be the $r$th order statistics.  It is shown \cite{caraux_bounds_1992,rychlik_stochastically_1992} that: 
\begin{equation}
P(d_{(r)}>t) \leq \min\left(1, S(t){J\over{J-r+1}}\right), \mbox{for } r = 1,\ldots, J. 
\label{eqn:tbound}
\end{equation}
This formula has been used to give an upper bound for Bayesian model checking with pivotal discrepancy measures calculated with posterior samples \cite{johnson_bayesian_2007,yuan_goodness--fit_2012}, which are correlated random variables. Here we apply this upper bound to a different scenario.  Suppose $p_{1},\ldots,p_{J}$ are replicated NRSP test p-values for a fitted model with the same dataset. We have shown that each $p_{j}$ is uniformly distributed on $(0,1)$ under the true model.  However, $p_{1},\ldots,p_{J}$  are correlated because they use the same dataset.  Applying the formula \eqref{eqn:tbound} to $d_{i}=-p_{i}$, we obtain the following inequality for the $r$th order statistics $p_{(r)}$: 
\begin{equation}
P(p_{(r)}<t) \leq \min\left(1,t{J\over r}\right). 
\label{eqn:pbound}
\end{equation}
Based on \eqref{eqn:pbound}, a p-value upper bound  for observed (simulated) $r$th statistics $p_{(r)}\obs$ is given by  $\min\left(1,p_{(r)}\obs{J\over r}\right)$.  To avoid the selection of $r$, we report the minimal upper bound for $r=1,\ldots, J$, denoted by $\pmin$:
\begin{equation}
\pmin = \min_{r=1,\ldots,J} \min\left(1,p_{(r)}\obs{J\over r}\right).
\end{equation}
The $\pmin$ is rather conservative for assessing model GOF because of its generality. When a model has a small $\pmin$, it is highly suspected that the model can be improved for better fitting the dataset. Considering the conservatism of $\pmin$, a rule of thumb for declaring model failure in practice should be much  larger (say 0.25 as suggested by \cite{yuan_goodness--fit_2012}) than the conventional $0.05$ for exact p-values. 

\section{Simulation Studies}\label{sec:sim}

\subsection{Detection of Mis-specified Distribution Family}\label{sec:wf}

In this simulation setting, we generated original failure times from a Weibull AFT model with $\log(T_{i}^{*})=2+x_{i}+\epsilon_{i}$, with $\epsilon_{i}$ generated from the extreme-value distribution with a shape parameter of 2, and censoring times $C_{i}$ were generated from $\exp(\theta)$. We set four different values of $\theta$ to obtain four different censoring rates ($c$): $0\%, 20\%, 50\%$, and $80\%$. We generated datasets with varying sample size $n$ ranging from 100 to 800. We then examined the performances of various residuals after we fitted the true model (Weibull AFT) and a wrong model (log-normal AFT) with the same linear link function, $\log(T_{i}^{*})=\beta_{0}+\beta_{1}x_{i}+\epsilon_{i}$, to these datasets. 

\begin{figure}[htbp]
 \centering
 \includegraphics[width=0.8\textwidth, height=2.5in]{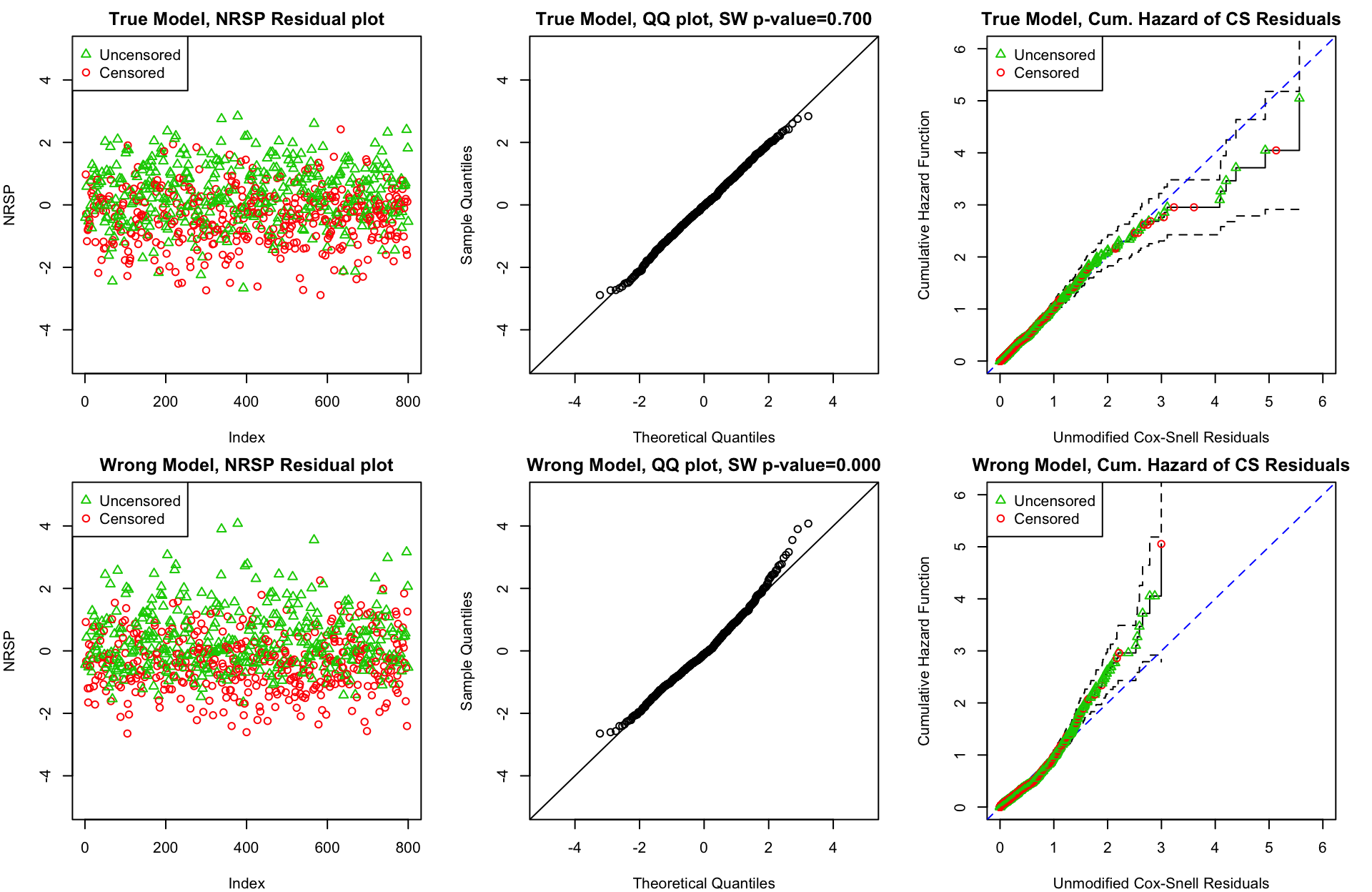} 
\caption{Performance of using the NRSP residuals as a graphical tool for detecting mis-specification of distribution family. The dataset has a sample size n = 800 and a censoring rate $c \approx 50\%$. The true model is a Weibull AFT, and the wrong model is a log-normal AFT.  Note that large NRSP residuals correspond to small failure times. An animated display of this figure with multiple simulated datasets is shown in  the URL given in Section \ref{sec:animation}.\label{fig:wf}
}
 \end{figure}

We first look at the performances of NRSP residuals on a single dataset with the sample size $n = 800$ and $\theta= 0.08$, which induce a censoring rate $c\approx 50\%$. Fig. \ref{fig:wf} displays the NRSP residuals against the index and their normal QQ plots under the true and the wrong models. Under the true model, the NRSP residuals are randomly scattered without exhibiting any pattern. They are mostly within the interval (-3, 3). The QQ plot of the NRSP residuals aligns nearly perfectly with the $45^{\circ}$ straight line. Under the wrong model,  the NRSP residuals are skewed to the right, and the QQ plot also indicates the deviation of NRSP residuals from the normal. Note that large NRSP residuals correspond to small failure times because of the descent of survival function. The scatterplot and QQ plot of NRSP residuals under the wrong log-normal model indicate that the true model for the dataset has more probability on the left than the fitted model (log-normal).  The corresponding residual and QQ plots of the NMSP and deviance residuals are given in Figure \ref{fig:suppresid-wf}. We see that the distributions of the NMSP and deviance residuals under the true model are far from the standard normal.  From looking at the CHF  of the wrong model (log-normal), we can draw the same conclusion regarding the problem of the model. However, we notice that the estimated cumulative hazard curve under the true model is not very straight, although it seems to be within the confidence bands. An animated display of Fig. \ref{fig:wf} with multiple simulated datasets is shown in the URL given in Section \ref{sec:animation}.

For demonstrating the performances of NRSP residuals in discriminating good and bad models with replicated simulated datasets, we evaluated the performances of a set of GOF tests applied to NRSP and other residuals with simulated datasets. The GOF test methods are given in the 2nd row of Table \ref{tab:wf}. The names of GOF test methods are denoted by ``R-T'' with ``R'' denoting residual name and ``T'' denoting test method. In particular, NUSP-CSF is the method that an extension of SF normality test is applied to NUSP residuals, which is implemented with \texttt{gofTestCensored} in R package \texttt{EnvStats}  \cite{millard_envstats_2018-1, steven_p_millard_author_envstats_2013}.  We generated 2000 datasets with the Weibull AFT model for each combination of a sample size $n$ and a censoring rate $c$, controlled by the mean of exponential censoring times. Using 1000 datasets generated under each scenario, we estimated the probabilities of model rejections when cutting GOF test p-values with 0.05. Table \ref{tab:wf} displays the percentages of model rejections of each GOF testing method under the true model and the wrong log-normal model. We see that the false-positive rates of NRSP-SW and NRSP-SF are close to the nominal level 0.05 for all scenarios (well-calibrated) and have good powers (true-positive rates) in detecting the incorrect choice of distribution family. NUSP-CSF has higher true-positive rates and lower false-positive rates than NRSP-SW and NRSP-SF. That is, NUSP-CSF is more discriminative than NRSP-SW and NRSP-SF. However, NRSP residuals plots can show more details about how a model misfits a dataset; for example, which observations are not accommodated by the model (i.e., outlier or divergent observations).

\begin{table}[htp]
\centering
\centering
\caption{Comparison of the percentages of model rejections of various GOF tests. A model is rejected when the test p-value is smaller than 0.05. Note that we use a random NRSP test p-value rather than the $\pmin$. The response variable is simulated from a Weibull AFT model with varying sample size and censoring rate. The wrong model is a log-normal AFT model with the same linear link function as the true model. \label{tab:wf}}
\resizebox{0.8\textwidth}{!}{
\begin{tabular}{rrrrrrrr|rrrrrr}
\hline
 & &\multicolumn{6}{c}{Under the true model} & \multicolumn{6}{c}{Under the wrong model} \\
 \hline
$n$&$100c$&{\scriptsize  NRSP-SW} &{\scriptsize  NRSP-SF}&{\scriptsize NUSP-CSF} &{\scriptsize  NRSP-KS}&{\scriptsize  NMSP-SW}&{\scriptsize  Dev-SW}&{\scriptsize  NRSP-SW}&{\scriptsize NRSP-SF}&{\scriptsize NUSP-CSF}&{\scriptsize  NRSP-KS}&{\scriptsize  NMSP-SW}&{\scriptsize  Dev-SW}\\
 \hline
100 & 0 & 4.40 & 4.95 & 4.95 & 0.05 & 4.40 & 4.20 & 94.10 & 93.55 & 93.55 & 10.05 & 94.10 & 93.25 \\ 
  200 & 0 & 3.35 & 3.70 & 3.70 & 0.00 & 3.35 & 3.10 & 99.85 & 99.75 & 99.75 & 42.05 & 99.85 & 99.85 \\ 
  400 & 0 & 4.40 & 4.55 & 4.55 & 0.00 & 4.40 & 4.60 & 100.00 & 100.00 & 100.00 & 90.30 & 100.00 & 100.00 \\ 
  800 & 0 & 5.10 & 5.15 & 5.15 & 0.00 & 5.10 & 5.60 & 100.00 & 100.00 & 100.00 & 99.95 & 100.00 & 100.00 \\ 
  \hline
  100 & 20 & 4.70 & 4.45 & 4.25 & 0.10 & 32.93 & 12.06 & 77.88 & 78.58 & 85.19 & 4.90 & 98.55 & 84.83 \\ 
  200 & 20 & 4.75 & 5.25 & 4.60 & 0.20 & 60.43 & 21.71 & 97.50 & 97.40 & 99.10 & 20.01 & 99.95 & 99.15 \\ 
  400 & 20 & 4.80 & 4.25 & 3.90 & 0.00 & 89.65 & 38.55 & 100.00 & 100.00 & 100.00 & 61.75 & 100.00 & 100.00 \\ 
  800 & 20 & 4.45 & 4.45 & 4.65 & 0.05 & 100.00 & 71.45 & 100.00 & 100.00 & 100.00 & 96.90 & 100.00 & 100.00 \\ 
  \hline
  100 & 50 & 4.13 & 4.58 & 2.82 & 0.35 & 99.90 & 89.28 & 44.34 & 49.57 & 60.34 & 2.37 & 100.00 & 98.74 \\ 
  200 & 50 & 4.37 & 4.37 & 2.26 & 0.75 & 100.00 & 99.55 & 75.25 & 78.16 & 90.16 & 7.93 & 100.00 & 100.00 \\ 
  400 & 50 & 4.94 & 4.94 & 2.47 & 0.55 & 100.00 & 100.00 & 96.92 & 97.43 & 99.65 & 19.93 & 100.00 & 100.00 \\ 
  800 & 50 & 4.87 & 4.67 & 2.16 & 0.40 & 100.00 & 100.00 & 100.00 & 100.00 & 100.00 & 55.49 & 100.00 & 100.00 \\ 
  \hline
  100 & 80 & 4.41 & 4.31 & 1.85 & 2.35 & 100.00 & 100.00 & 11.62 & 15.37 & 22.33 & 2.15 & 100.00 & 100.00 \\ 
  200 & 80 & 4.31 & 4.81 & 2.06 & 1.95 & 100.00 & 100.00 & 23.01 & 29.22 & 45.96 & 3.46 & 100.00 & 100.00 \\ 
  400 & 80 & 4.71 & 4.31 & 1.10 & 2.20 & 100.00 & 100.00 & 46.97 & 52.33 & 75.71 & 3.46 & 100.00 & 100.00 \\ 
  800 & 80 & 5.27 & 5.17 & 0.80 & 2.26 & 100.00 & 100.00 & 77.17 & 80.98 & 96.54 & 6.12 & 100.00 & 100.00 \\ 
  \hline
\end{tabular}
}
\end{table}
Table \ref{tab:wf} also shows that the performances of other GOF tests are not satisfactory. Because NMSP and deviance residuals are not normally distributed when $c>0$, NMSP-SW and Dev-SW have very high false-positive rates when $c>0$. The true-positive rates of the NMSP-SW and Dev-SW methods are very high. However, the high true-positive rates of NMSP-SW and Dev-SW do not imply that they are discriminative because they have very high false-positive rates (nearly 100\% when $c$ is large).  NRSP-KS has low false-positive and true positive rates because the KS test is more affected by the double use of data in estimating parameters and calculating residuals than SW and SF tests; see a more dedicated study in appended Section \ref{sec:ks}. In appended Fig. \ref{fig:supphist-wf}, we provide the histograms of test p-values of NRSP-SW, NMSP-SW and Dev-SW when $n=800$ and $c\approx50\%$ for showing more details of their performances. 
 
\subsection{Detection of  Non-linear Covariate Effect}\label{sec:nl}
In this section, we demonstrate the effectiveness of the NRSP residuals in detecting non-linear covariate effects. The response variable is simulated from a Weibull AFT regression model with a non-linear link function: $\log (T_i^{*})= 2+5\sin(2x_{i}) + \epsilon_i$. The covariate $x_{i}$ was generated uniformly on $(0, {3{\pi}/2})$.  The shape parameter of the Weibull distribution was set as 1.8. The censoring times $C_{i}$ were generated from $\exp(\theta)$ with $\theta$ varied for obtaining different censoring rates. We considered fitting a Weibull AFT model assuming $\log (T_i^{*})= \beta_{0}+\beta_{1}x_{i} + \epsilon_i$ as a wrong model, and fitting a Weibull AFT model assuming $\log(T_{i}^{*})=\beta_{0}+\beta_{1}\sin(2x_{i})+\epsilon_{i}$ as the true model. 

\begin{figure}[htp]
 \centering
   \includegraphics[width=0.8\textwidth, height=2.5in]{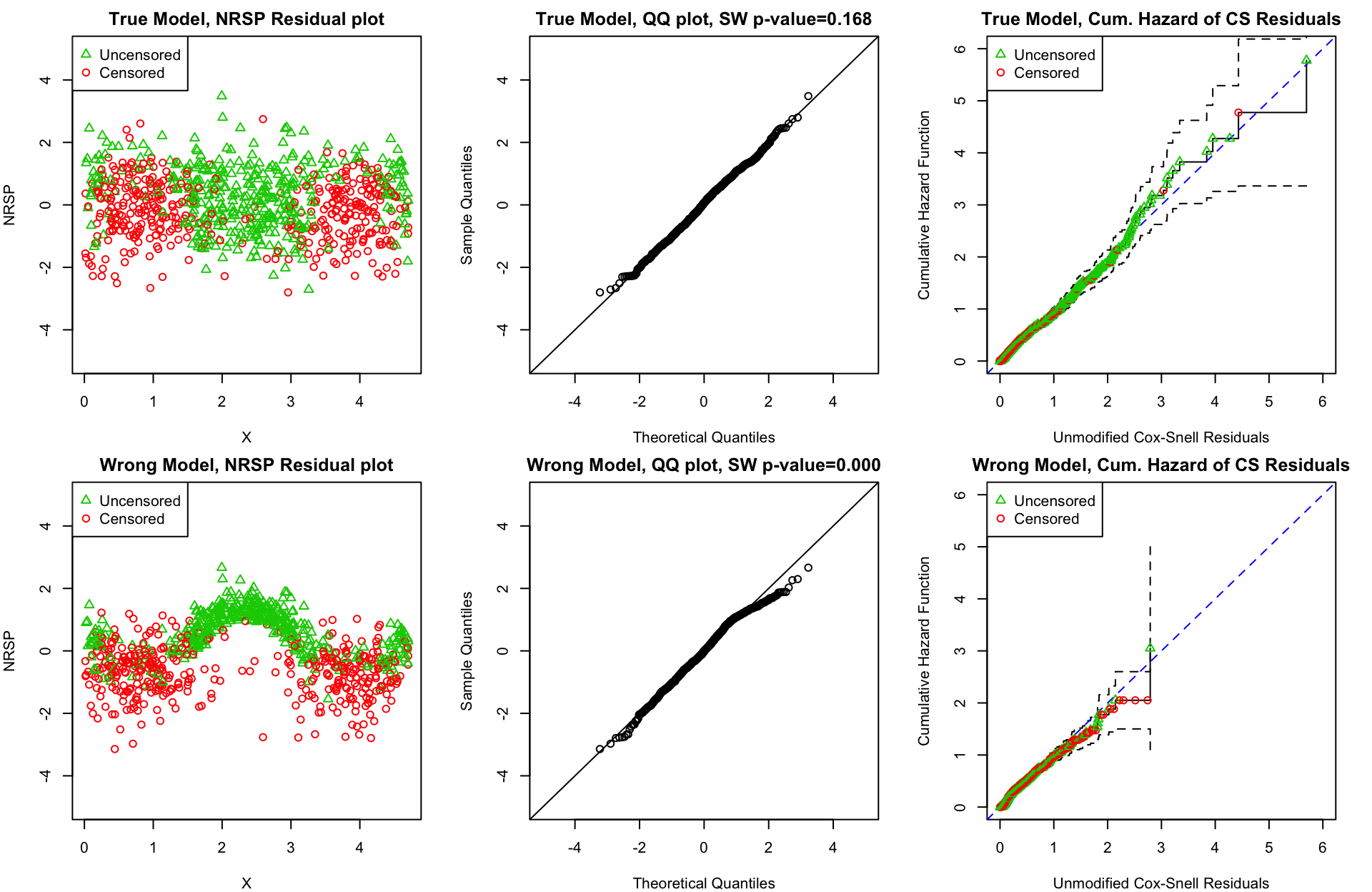} 

\caption{Performance of the NRSP residuals as a graphical tool for detecting non-linear effect in covariate. The dataset has a sample size n = 800 and a censoring rate $c \approx 50\%$. The true model is a Weibull AFT model $\log (T_i^{*})= \beta_{0}+\beta_{1}\sin(2x_i) + \epsilon_i$ and the wrong model is a Weibull AFT model $\log (T_i^{*})= \beta_{0}+\beta_{1}x_{i} + \epsilon_i$. Note that large NRSP residuals correspond to small failure times.  An animated display of this figure with multiple simulated datasets is shown in  the URL given in Section \ref{sec:animation}. \label{fig:RSP2}}
 \end{figure}

We first look at the performances of NRSP residuals on a single dataset with the sample size $n = 800$ and $c\approx 50\%$. Figure \ref{fig:RSP2} displays the NRSP residuals against the covariate $x_{i}$ and their normal QQ plots. Under the true model, the residuals are mostly bounded between -3 and 3 as standard normal deviates without a visible pattern; the QQ plot aligns well with the $45^{\circ}$ straight line. Under the wrong model, a non-linear pattern in the NRSP residual scatterplot is obvious, and the QQ plot deviates from the $45^{\circ}$ straight line. The CHF of CS residuals under the wrong model aligns well with the $45^{\circ}$ straight line. Therefore, for this example, the visual inspection of the CHF of CS residuals fails to detect the non-linearity in the dataset. The scatterplots and QQ plots of the NMSP and deviance residuals are given in Figure \ref{fig:NCS2}. There are non-linear patterns in their scatterplots under the true and wrong models; hence, we cannot use the non-linear patterns to distinguish good and bad models. 

We used simulated datasets to evaluate the performances of a set of statistical tests applied to NRSP and other residuals.  GOF tests only compare the distribution of residuals as univariate data with a hypothetical distribution; hence, they may not detect implausible assumptions in linking $\Tistar$ and $x_{i}$. Therefore, we also investigated a non-linearity test with NRSP residuals, denoted by \textbf{NRSP-AOV}.  In this test, we first divide NRSP residuals into $k=10$ groups by cutting the fitted values into equally-spaced intervals; then, we apply the $F$-test in ANOVA to test whether the means of NRSP residuals are equal in the $k$ groups. We generated 2000 datasets for each combination of a sample size $n$ and a censoring rate $c$ from the true model for estimating the percentages of model rejections (i.e.,  test p-values $< 0.05$) of each test method.  The results are shown in Table \ref{tab:powersw2}.  NRSP-SW and NRSP-SF methods have false-positive rates close to the nominal level 0.05 for all scenarios and good power in detecting non-linearity. The performances of NMSP-SW, Dev-SW, and NRSP-KS are not satisfactory, as described in Section \ref{sec:wf}. We see that NUSP-CSF is more discriminative than NRSP-SW and NRSP-SF. However,  the NRSP residuals enable us to conduct non-linearity diagnostics in addition to the GOF checking. Table \ref{tab:powersw2} shows that the NRSP-AOV can detect the non-linearity with very high powers (nearly 100\%), significantly higher than those of the GOF tests, including NUSP-CSF.

\begin{table}[htbp]
\centering
\centering
\caption{Comparison of the percentages of model rejections of various statistical tests. A model is rejected when the test p-value is smaller than 0.05. Note that we use a random NRSP test p-value rather than the $\pmin$. The response variable is simulated from a Weibull AFT model $\log (T_i^{*})= 2+5 \sin(2x_i) + \epsilon_i$ with varying sample size and censoring rate. We consider fitting a Weibull AFT regression model $\log (T_i^{*})= \beta_{0}+\beta_{1}x_{i} + \epsilon_i$ as a wrong model. 
\label{tab:powersw2}}
\resizebox{\textwidth}{!}{
\begin{tabular}{rrrrrrrrr|rrr>{\bfseries}rrrr}
\hline
 & &\multicolumn{7}{c}{Under the true model} & \multicolumn{7}{c}{Under the wrong model} \\
 \hline
$n$&$100c$&{\scriptsize  NRSP-SW} &{\scriptsize NRSP-SF}&{\scriptsize NUSP-CSF}&{\scriptsize NRSP-AOV}&{\scriptsize  NRSP-KS}&{\scriptsize  NMSP-SW}&{\scriptsize  Dev-SW}&{\scriptsize  NRSP-SW}&{\scriptsize NRSP-SF}&{\scriptsize NUSP-CSF}&{\scriptsize NRSP-AOV}&{\scriptsize  NRSP-KS}&{\scriptsize  NMSP-SW}&{\scriptsize  Dev-SW}\\
 \hline
 100 & 0 & 4.90 & 4.65 & 4.65 & 3.00 & 0.00 & 4.90 & 4.80 & 62.30 & 43.85 & 43.85 & 100.00 & 0.40 & 62.30 & 63.85 \\ 
  200 & 0 & 3.75 & 4.60 & 4.60 & 3.75 & 0.00 & 3.75 & 3.75 & 95.00 & 89.85 & 89.85 & 100.00 & 5.90 & 95.00 & 95.75 \\ 
  400 & 0 & 4.50 & 4.15 & 4.15 & 3.60 & 0.00 & 4.50 & 4.20 & 99.95 & 99.95 & 99.95 & 100.00 & 45.80 & 99.95 & 99.95 \\ 
  800 & 0 & 4.45 & 4.50 & 4.50 & 3.05 & 0.05 & 4.45 & 5.10 & 100.00 & 100.00 & 100.00 & 100.00 & 98.40 & 100.00 & 100.00 \\ 
  \hline
  100 & 20 & 4.90 & 5.10 & 4.80 & 3.50 & 0.10 & 30.35 & 12.40 & 50.55 & 34.60 & 50.70 & 100.00 & 0.20 & 78.15 & 80.00 \\ 
  200 & 20 & 5.45 & 5.15 & 5.35 & 4.35 & 0.00 & 55.20 & 20.25 & 88.05 & 80.05 & 91.55 & 100.00 & 2.00 & 98.85 & 98.85 \\ 
  400 & 20 & 5.00 & 5.25 & 4.45 & 3.30 & 0.05 & 88.00 & 37.90 & 99.75 & 99.60 & 99.90 & 100.00 & 25.20 & 100.00 & 100.00 \\ 
  800 & 20 & 5.35 & 5.60 & 5.00 & 2.60 & 0.20 & 99.60 & 68.80 & 100.00 & 100.00 & 100.00 & 100.00 & 89.25 & 100.00 & 100.00 \\ 
   \hline
  100 & 50 & 4.45 & 4.95 & 3.25 & 3.75 & 0.35 & 99.95 & 94.25 & 40.35 & 26.55 & 56.95 & 100.00 & 0.30 & 99.80 & 99.80 \\ 
  200 & 50 & 5.70 & 6.30 & 2.65 & 3.40 & 0.55 & 100.00 & 99.75 & 82.35 & 72.05 & 92.80 & 100.00 & 0.60 & 100.00 & 100.00 \\ 
  400 & 50 & 5.45 & 5.15 & 1.60 & 3.20 & 0.85 & 100.00 & 100.00 & 99.50 & 99.05 & 99.85 & 100.00 & 7.95 & 100.00 & 100.00 \\ 
  800 & 50 & 4.55 & 4.10 & 1.35 & 3.60 & 0.65 & 100.00 & 100.00 & 100.00 & 100.00 & 100.00 & 100.00 & 55.75 & 100.00 & 100.00 \\ 
   \hline
  100 & 80 & 4.46 & 4.67 & 1.73 & 2.64 & 1.98 & 100.00 & 100.00 & 8.52 & 5.33 & 21.36 & 92.03 & 1.12 & 100.00 & 100.00 \\ 
  200 & 80 & 4.28 & 4.58 & 1.83 & 3.16 & 2.04 & 100.00 & 100.00 & 24.49 & 14.77 & 47.00 & 99.90 & 2.49 & 100.00 & 100.00 \\ 
  400 & 80 & 5.07 & 5.23 & 0.72 & 4.20 & 3.02 & 100.00 & 100.00 & 59.92 & 46.44 & 83.24 & 100.00 & 2.20 & 100.00 & 100.00 \\ 
  800 & 80 & 4.90 & 5.01 & 0.46 & 3.41 & 2.17 & 100.00 & 100.00 & 93.86 & 89.73 & 98.61 & 100.00 & 4.44 & 100.00 & 100.00 \\  \hline
\end{tabular}
}
\end{table}

We also evaluated the performances of NRSP model diagnostics when we reject models with $\pmin\leq 0.05$; see the appended Table \ref{tab:powersw_pmin}. As expected, the powers of NRSP tests with $\pmin\leq 0.05$ become smaller than those of NRSP tests with random p-values because $\pmin$ is a p-value upper bound.  However, the powers of NRSP-AOV with $\pmin\leq 0.05$ are still near $100\%$ for this example, although $\pmin$ is generally conservative. 

\section{A Real Data Example}\label{sec:realdata}
This section will demonstrate the power of the model diagnostics with NRSP residuals using a recurrence-free times dataset of breast cancer patients \cite{schumacher_randomized_1994, sauerbrei_building_1999}. A cohort study of breast cancer in a large number of hospitals was carried out by the German Breast Cancer Study Group to compare three cycles of chemotherapy with six cycles and also to investigate the effect of additional hormonal treatment consisting of a daily dose of 30 mg of tamoxifen over two years. The patients in the study had primary histologically proven non-metastatic node-positive breast cancer who had been treated with mastectomy. The dataset is consolidated from 41 centres with a total of 686 patients. The censoring rate is 56.5\%. The response variable of interest is the recurrence-free time, which is the time from entry to the study until a recurrence of cancer or death. We consider the following covariates: the tamoxifen treatment indicator, patient age, menopausal status, size and grade of the tumour, number of positive lymph nodes, progesterone and estrogen receptor status. More descriptions of these variables can be found from  \cite{schumacher_randomized_1994} and Table \ref{tab:variable} in the appendix.
 
\begin{figure}[htp]
 \centering
\includegraphics[width=\textwidth, height=4in]{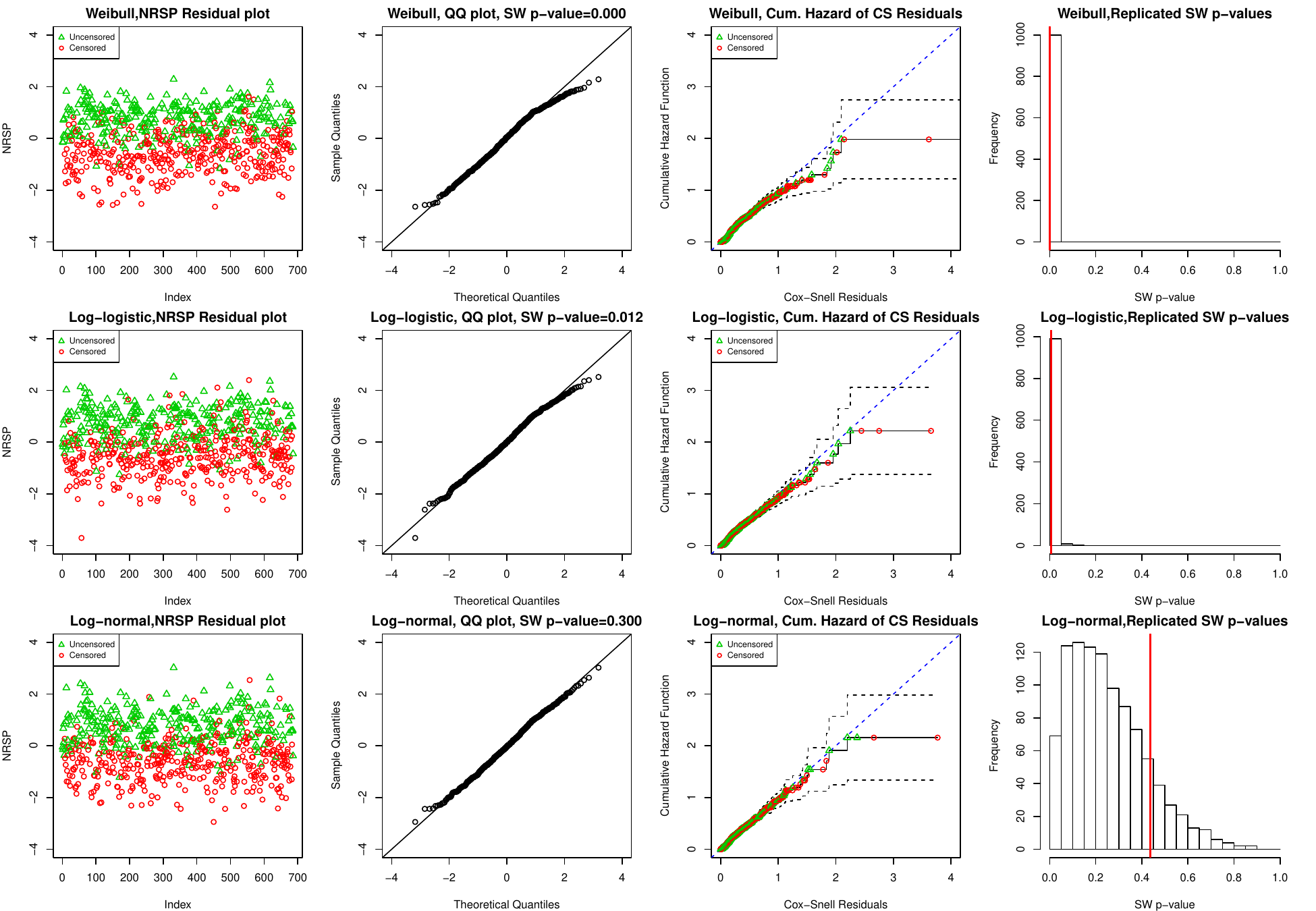}
\caption{NRSP residuals of the Weibull, log-logistic, and log-normal AFT models fitted to the breast cancer patients dataset.  The last column presents the histograms of 1000 replicated NRSP-SW p-values of each model. The vertical red lines indicate $\pmin$ calculated with the 1000 replicated NRSP p-values . An animated display of this figure for replicated NRSP residuals except the last column is shown in  the URL given in Section \ref{sec:animation}.\label{fig:realdata}}
 \end{figure}

We fitted Weibull, log-logistic and log-normal AFT models with the available variables to the recurrence-free failure times. Table \ref{tab:covariates} (in the appendix) shows the estimated regression coefficients, the corresponding standard errors and p-values for the covariate effects from fitting the three AFT models.  The third column of Figure \ref{fig:realdata} displays the estimated CHFs of CS residuals of the three models. All of the three curves appear to align well with the $45^{\circ}$ straight line. The confidence bands estimated with at least one uncensored observation contain the $45^{\circ}$ straight line. However, we will show that all of the three models misfit the data with model diagnostics based on NRSP and NUSP residuals.

We calculated the NRSP residuals of the three AFT models. The first and second columns of Figure \ref{fig:realdata} present the scatterplots and QQ plots of the NRSP residuals versus the index for each model.  For the Weibull and log-logistic models, their NRSP residuals skew to the left; the QQ plots of the NRSP residuals also deviate from the $45^{\circ}$ straight line in the upper tail.  These observations suggest that a more appropriate model for the dataset should assign more probability to the right than Weibull and log-logistic models (note that small NRSP residuals correspond to large $\Tistar$). In contrast, for the log-normal model, the NRSP residuals are mostly between -3 and 3 and do not exhibit a visible pattern; the QQ plot of NRSP residuals aligns well with the  $45^{\circ}$ straight line. 

We further conducted GOF tests for the three models. The NUSP-CSF p-values for the three models are given in the 2nd row of Table \ref{tab:realdata}, which shows that the Weibull and log-logistic models do not fit the data well, and the log-normal model appears a good fit. One difficulty in applying NRSP test methods is the fluctuation in test p-values due to the randomness in generating NRSP residuals. One way to remedy is to generate multiple sets of NRSP residuals. We then look at the histograms of the replicated test p-values and calculate the p-value upper bound $\pmin$ as described in Section \ref{sec:pmin}. We generated 1000 realizations of the NRSP residuals for this dataset,  and consequently, we obtained 1000 replicated NRSP-SW p-values for each model. We show an animated display of the scatterplot and QQ plots of the replicated NRSP residuals in the URL given in Section \ref{sec:animation}. From the animation, we see that we can identify the misfits of Weibull and log-logistic models in most sets of replicated residuals. The fourth column of Figure \ref{fig:realdata} displays the histograms of 1000 replicated NRSP-SW test p-values. The $\pmin$ and the percentages of replicated NRSP-SW and NRSP-SF p-values being $< 0.05$ for each model are given in Table \ref{tab:realdata}.  The $\pmin$ values of the Weibull and log-logistic models are also significantly smaller than 0.05. The small $\pmin$ values provide strong evidence that the Weibull and log-logistic models do not fit the dataset well, but the log-normal seems a good model for the dataset with these GOF tests.

\begin{table}[htp]
\centering
\caption{Model diagnostic test p-values or $\pmin$ for NRSP tests and AIC values of different models for the breast cancer data. The numbers in brackets for NRSP tests are the percentages of replicated NRSP test p-values being $\leq 0.05$.  \label{tab:realdata}}
\begin{tabular}{l|r|r|r|r}
\hline
Model&Weibull & log-logistic  & log-normal&log-normal with log(nodes)\\ 

\hline
AIC &5181 & 5153 & 5139& 5121\\
\hline
 NUSP-CSF p-values & 1.69e-5  & 1.94e-3& 0.133& 0.172 \\
\hline
NRSP-SW $\pmin$ &2.43e-05 (100.00) & 6.01e-03 (99.00) & 4.36e-01 (\ \ 6.90) & 5.29e-01 (5.70) \\
\hline
NRSP-SF $\pmin$ & 9.50e-05 (100.00) & 1.52e-02 (97.00) & 4.85e-01 (\ \ 5.80) & 6.37e-01 (3.50) \\
\hline
NRSP-AOV $\pmin$ & 1.97e-02 (\ \ 60.40) & 7.21e-02 (46.00) & \textbf{5.22e-02 (52.40)} & 9.99e-01 (0.50) \\ 
\hline

\end{tabular}
\end{table}
 
 We also calculated the NMSP and deviance residuals for the three models. Figure \ref{fig:NCS_WB_LN_LL} and \ref{fig:DEV_WB_LN_LL} in the appendix display the residual plots and the QQ plots of the NMSP and deviance residuals. We see that all of these residuals deviate from a normal distribution due to censoring. Therefore, the NMSP and deviance residuals fail to distinguish the GOFs of these three models to this dataset.

We also compared the model checking results based on the NRSP residuals to the model comparison results based on AICs. First, let us clarify the difference between model \textit{checking} and model \textit{comparison}. AIC is a measure of the out-of-sample predictive performance of a model. Even when none of the models in a set fit a dataset well, AIC will always choose one as the best model. Therefore, the model comparison alone is insufficient for model evaluation. We still need to conduct model diagnostics for the best model chosen by AIC or other information criteria. On the other hand, the model with better AIC is believed to fit the dataset better. Therefore, it is still meaningful to compare the model comparison results based on AIC with the model diagnostics results. Table \ref{tab:realdata} displays the AIC values of the three fitted AFT models. We see that the log-normal model has a lower AIC than the Weibull and log-logistic models. 

\begin{figure}[htbp]
\begin{center}
\includegraphics[width=0.8\textwidth, height=0.45\textwidth]{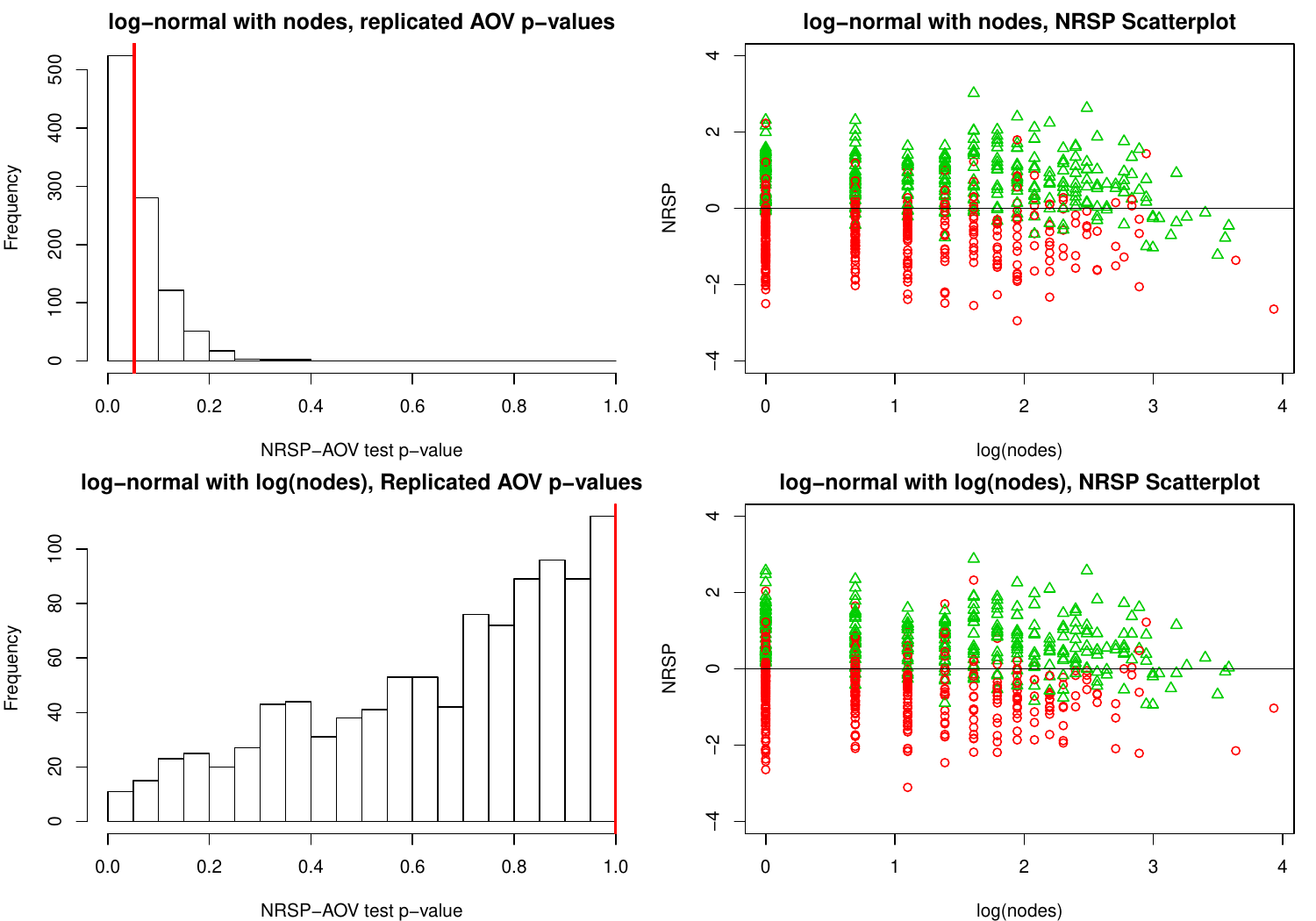}
\caption{NRSP non-linearity residual diagnosis for the breast cancer data. The red vertical lines in the histograms of replicated NRSP-AOV p-values show the  $\pmin$ values.}
\label{fig:realdataNL}
\end{center}
\end{figure}

The GOF tests with NRSP residuals and NUSP-CSF report fairly large p-values for the log-normal model. Despite these GOF test results, we further checked whether the linear assumption is plausible with the NRSP-AOV method (\textit{described in Section \ref{sec:nl}}). We calculated 1000 replicated NRSP-AOV p-values.  The top-left plot of  Figure \ref{fig:realdataNL} shows the histogram of these replicated NRSP-AOV p-values for the log-normal model. The percentage of NRSP-AOV  p-values $< 0.05$ is about $50\%$, and the $\pmin$ is only slightly larger than 0.05. Therefore, we have fair evidence to suspect that there is non-linearity. We drew the NRSP residuals against each covariate and linear predictor, one of which is shown in the top-right plot of Fig. \ref{fig:realdataNL}. This plot shows that large ``nodes'' values appear to have small NRSP residuals (not symmetric about 0). We tried a logarithm transformation for nodes and fitted a log-normal model with log(nodes) and other variables. This model is labelled as ``log-normal with log(nodes)'' in Table \ref{tab:realdata} and Fig. \ref{fig:realdataNL}. We see that the NRSP residuals of this model are more homogeneous along with the variable ``nodes'', as shown in the bottom-right plot of Fig. \ref{fig:realdataNL}. The replicated NRSP-AOV p-values (bottom-left plot of Fig. \ref{fig:realdataNL}) are mostly larger than 0.05 with $\pmin$ near 1. Furthermore, the AIC value of this model, as shown in the last column of Table \ref{tab:realdata}, supports that the log transformation results in a better model.  This above analysis clearly shows that the non-linearity diagnostics with NRSP residuals successfully detect the non-linear effect of ``nodes'' in this dataset. However, the effect is too subtle to be detected by the above GOF tests, including the NUSP-CSF test.

\section{Conclusions and Discussions}\label{sec:conc}
This paper has proposed using randomized survival probabilities (RSPs) to conduct model diagnostics for censored regression. We have proved that RSPs always have the uniform distribution on $(0,1)$ under the true model. Consequently, NRSP residuals are approximately distributed with $N(0,1)$ under the true model. With this unified reference distribution for NRSP residuals, we can conduct a wide variety of residual diagnostics for censored regression. Our simulation studies show that, although the GOF tests with NRSP residuals are not as powerful as a traditional GOF test method, a non-linearity test with NRSP residuals has significantly higher power in detecting non-linearity.  The real data analysis shows that the NRSP residual diagnostics successfully captures a subtle non-linear relationship in the dataset, which is not detected by the graphical diagnostics with the CS residuals and existing GOF tests. 

The NRSP residual for one-sided censored regression, as described in this article, can be easily extended to interval-censored regression by drawing a random number between the two survival probabilities calculated on the two bounds of each interval.  The NRSP residual for interval-censored regression can be regarded  as an extension of the randomized quantile residual \cite{dunn_randomized_1996} for count regression if we consider a count observation as an observation of a continuous variable censored by fixed integer intervals.  

To overcome the randomness in NRSP testing p-values, we need to devise valid methods to obtain non-random GOF test p-values based on NRSP residuals. In this article, we describe a method for obtaining a p-value upper bound $\pmin$. $\pmin$ is informative when the model departure is clear. However, it is generally conservative. We desire to have a more powerful and non-random summary of the replicated NRSP test p-values.  Our preliminary results (not shown in this article) show that averaging replicated NRSP test p-values can boost the discriminative power of NRSP-SW and NRSP-SF methods to the power of NUSP-CSF, as measured by ROC curves. However,  the averages of replicated  NRSP-SW or NRSP-SF  p-values are no longer uniformly distributed. We believe that to characterize the distribution of the average of replicated NRSP testing p-values or to obtain a rule of thumb is an interesting topic. 

This article shows that a simple non-linearity test by applying ANOVA to NRSP residuals has superior power than GOF tests in detecting non-linearity. We expect that many other specific model mis-specification tests that target a particular model discrepancy have higher powers than GOF tests. For example, statistical tests for checking proportional hazard assumption in Cox regression seem to be demanded very often; see \cite{grambsch_proportional_1994} among others. We believe that developing specific quantitative and graphical diagnostic tools based on NRSP  residuals will be fruitful because of the explicit characterization of the approximate standard normal distribution of NRSPs under the true model.

\appendix


\vspace*{-15pt}

\section{Additional Figures and Tables}
Additional figures, tables, and simulation results can be found from \url{https://onlinelibrary.wiley.com/action/downloadSupplement?doi=10.1002%2Fsim.8852&file=sim8852-sup-0001-supinfo.pdf}.

\vspace*{-15pt}

\section{R Functions, Demonstration Examples, and Data Availability}\label{sec:animation}
R functions for computing NRSP residuals for \texttt{survreg} and \texttt{coxph} objects with demonstration examples  and the dataset used in this paper are available here:  \url{https://longhaisk.github.io/software/NRSP/}.  

\vspace*{-15pt}

\section*{Acknowledgement} We much appreciate the anonymous referees of \textit{Statistics in Medicine} for their useful comments to improve this paper. 

\vspace*{-15pt}

\bibliographystyle{WileyNJD-AMA}
\bibliography{ModelChecking,sim}

 \end{document}